\documentclass[twocolumn,prl,amsmath,amssymb,8pt,showpacs]{revtex4}
\usepackage{dcolumn}% Align table columns on decimal point
\usepackage{bm}% bold math
\usepackage{graphics}
\usepackage{graphicx}
\usepackage{amssymb}
\usepackage{amsmath}
\usepackage {longtable}
\allowdisplaybreaks

\begin{document}

\title{A MAGNETIC-FIELD-EFFECT TRANSISTOR AND SPIN TRANSPORT}

\author{R.N.Gurzhi}
\affiliation{B.Verkin Institute for Low Temperature Physics \&
Engineering of the National Academy of Sciences of the Ukraine, 47
Lenin Ave, Kharkov, 61103, Ukraine}
\author{A.N.Kalinenko}
\affiliation{B.Verkin Institute for Low Temperature Physics \&
Engineering of the National Academy of Sciences of the Ukraine, 47
Lenin Ave, Kharkov, 61103, Ukraine}
\author{A.I.Kopeliovich}
\affiliation{B.Verkin Institute for Low Temperature Physics \&
Engineering of the National Academy of Sciences of the Ukraine, 47
Lenin Ave, Kharkov, 61103, Ukraine}
\author{A.V.Yanovsky}
\affiliation{B.Verkin Institute for Low Temperature Physics \&
Engineering of the National Academy of Sciences of the Ukraine, 47
Lenin Ave, Kharkov, 61103, Ukraine}
\author{E.N.Bogachek}
\affiliation{School of Physics, Georgia Institute of Technology,
Atlanta, GA 30332-0430, USA}
\author{Uzi Landman}
\affiliation{School of Physics, Georgia Institute of Technology,
Atlanta, GA 30332-0430, USA}

\begin{abstract}
A magnetic-field-effect transistor is proposed that generates a
spin-polarized current and exhibits a giant negative
magnetoresitance. The device consists of a nonmagnetic conducting
channel (wire or strip) wrapped, or sandwiched, by a grounded
magnetic shell. The process underlying the operation of the device
is the withdrawal of one of the spin components from the channel,
and its dissipation through the grounded boundaries of the
magnetic shell, resulting in a spin-polarized current in the
nonmagnetic channel. The device may generate an almost fully
spin-polarized current,  and a giant  negative magnetoresistance
effect is predicted.
\end{abstract}
%\date{}
\pacs{72.25.Hg, 72.25Mk, 73.40.Sx, 73.61.Ga}
 \maketitle

Spintronic devices for storage and transport of information and
for quantum computations have been the subject of
increasing interest \cite{Aws,Dat}.
%, both  because of the novel
%physics involved in such devices and due to their potential
%technological and economical importance \cite{Aws}. The key idea
%underlying research in this field is the use of the spin degree of
%freedom for data manipulations, in addition to and/or as an
%alternative to the electronic charge. Among the main current
%proposals for such devices are schemes and designs aiming at
%fabrication of new types of transistors, such as a spin
%field-effect transistor \cite{Dat}.
%\includegraphics[height=10cm,width=8cm]{fig1.bmp}
%
%
\begin{figure}
\includegraphics[height=4.5cm,width=8cm]{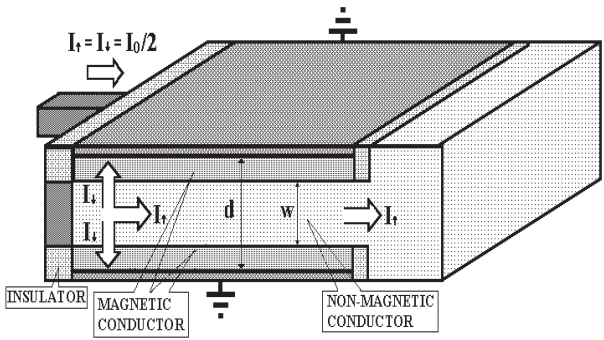}
\caption{A schematic  of the spin-guide. w is the width of the
nonmagnetic channel and d is the distance between the grounded
magnetic  contacts.} \label{fig1}
\end{figure}
Here we propose a scheme for a magnetic-field-effect transistor
(MFET), where the electric current flow is governed by a magnetic
field in contrast to conventional electronic devices (e.g. the
silicon-based FET) which are controlled by an
electric field.
The purpose of the proposed spin-guide scheme pertains to
generation and transport of electric currents
characterized by a high degree of spin-polarization.

We propose to use a ``sandwich'' configuration, with a nonmagnetic
(NM) conducting channel and a surrounding magnetic material (MM)
whose external boundaries are grounded (see Fig.~\ref{fig1}).
Electric current flows parallel to the NM/MM interface, instead of
being normal to it as in a spin-filter \cite{Aro,Son,Mol2}.
Underlying the operation of the device is the removal of one of
the current spin-components from the NM channel, whereas in the
spin-filter scheme spin polarization in a NM conductor is created
by electrons injected from a magnetic material.

Let an unpolarized constant current be driven through the channel
entrance.
Away from the channel entrance a
difference will develop between the spin-up and spin-down
currents. Non-equilibrium electrons with one of the spin
directions (that coincides with the majority-spin direction in the
magnetic layer, spin-down for example) will preferentially leave
the nonmagnetic channel - that is, the transparency of the NM/MM
interface is different for spin-up and spin-down electrons due to
the conductivity difference in these materials. With high
probability (particularly for a thin magnetic shell) these
electrons will dissipate at the grounded external boundary without
return to the channel. Consequently, a polarized electric current
is generated in the channel with the polarization increasing as a
function of the distance from the channel entrance (due to
depletion of the spin-down carriers).

For implementation of the device we consider two classes of
magnetic materials for the magnetic shell.
(1) Dilute II-VI magnetic semiconductors (DMS): these compounds
may have a sufficiently high degree of spin polarization
\cite{Mol2} because of the very large Zeeman splitting of the spin
subbands.
When the Fermi level in the DMS lies
below the bottom of one of the spin subbands, nearly full
spin-polarization of the DMS may be reached. Thus, using a DMS
(consisting e.g. of (Zn,Mn,Be)Se as the MM) and a lattice matched
(Zn,Be)Se as the NM channel, one may achieve conditions where all
the electrons in the magnetic material are fully spin-polarized
and the magnetic shell will not transmit electrons with one of the
spin directions (spin-up in our example, see Fig.{~\ref{fig1}}).
(2) Alternatively,
ferromagnetic metals (like Ni, Fe or Co) may be used for the
magnetic shell. In contrast to ordinary electronic devices where a
combination of a metal with a semiconductor is used, our scheme
may be implemented as an all-metal device
%; i.e.one may interface a NM semiconductor, or a NM metal, as the inner
%channel, with surrounding ferromagnetic metal layers.

The above device exhibits sensitivity to changes in the magnetic
field. Indeed, the aforementioned selective transparency of the
NM/MM interface provides  different decay length-scales for the
spin-up and spin-down electrons along the channel. Therefore, if
we create an unmagnetized magnetic shell by switching-off the
magnetic field, the dissipation of all the non-equilibrium
electrons (i.e. of both spin directions) at the grounded boundary
will be faster then the rate of their arrival to the channel exit,
i.e. almost all the electrons may not reach the exit of the spin
guide because of a larger probability to dissipate to the
grounding. Thus, by changing the applied magnetic field we may
change significantly the resistance of the spin-guide device (that
is, a strong negative magnetoresistance effect is predicted), and
it may operate as a transistor governed by the magnetic field,
i.e. a MFET.

%We turn now to calculation of the currents in the spin-guide
%scheme.
Consider $\mu_{\uparrow,\downarrow}$ - the non-equilibrium
parts of the electrochemical potentials for the two spin directions.
In the diffusive regime, the spin transport is described by
(see, e.g. \cite{Son})
\begin{equation}
\label{eq1}
div(\sigma_{ \uparrow , \downarrow } \nabla \mu _{
\uparrow , \downarrow } ) = \Pi _{0} e^{2}\tau_{sf}^{-1}(\mu _{
\uparrow , \downarrow } - \mu _{ \downarrow , \uparrow } ), \\
\end{equation}
where $\Pi_{0} ^{ - 1} = \Pi_{ \downarrow } ^{ - 1} + \Pi_{
\uparrow }^{-1}$, $\Pi_{\uparrow, \downarrow}  $ are the densities
of states at the Fermi level of the up and down spins, $\tau_{sf}$
is the spin-flip scattering time, and $\sigma_{\uparrow ,
\downarrow} $  are the corresponding conductivities.
Eqs.(\ref{eq1}) holds under the assumption that both the spin-flip
mean-free-paths $l^{sf}_{\uparrow, \downarrow}= v_{F \uparrow,
\downarrow} \tau_{sf}$ ($v_{F \uparrow, \downarrow}$ are the Fermi
velocities of the spin-up and spin-down electrons), and the widths
of the channel ($w$) and  the magnetic shell ($w_{M}=(d-w)/2$)
exceed significantly the diffusion step-lengths $l_{\uparrow,
\downarrow}$
i.e.
$l^{sf}_{\uparrow, \downarrow},w, w_{M} \gg
l_{\uparrow, \downarrow}$
; otherwise, the problem should be
treated within the kinetic equation approach. The electric current
densities $J_{\uparrow, \downarrow}=(-\sigma_{ \uparrow ,
\downarrow}/e)\nabla \mu_{ \uparrow , \downarrow }$ are related to
the electrochemical potentials via Ohm's law.

Let the $x$ axis lie in the middle of the channel and be directed
along it, and take the $z$ axis to be perpendicular to the interfacial
planes, with the origin of the coordinate system located in the
center of the entrance into the channel (see Fig.{\ref{fig1}}).
The conductivity of the NM channel is spin independent and
constant ($\sigma_{N \uparrow}=\sigma_{N \downarrow} =
\sigma_{N}$), and we impose the boundary conditions
$\mu_{\uparrow,\downarrow}=0$ at $z=\pm d/2$ reflecting the grounding
of the external planar boundaries.
%, and due to electric neutrality we
%obtain for chemical ($\eta$) and electrical ($\varphi$) potentials
%$\eta_{\uparrow,\downarrow}=\varphi=0$ at $z=\pm d/2$. It is
%reasonable to take the same potentials at the exit from the
%channel to avoid foreign currents.
The solutions of Eq.(\ref{eq1}) that we seek consist of a sum of
terms, with each expressed as products of two functions, one
depending on the x-variable and the other depending on the
z-coordinate and on the discrete indices ``$\pm$''
(see Eq. (\ref{eq4}) below). To find the general
solution of Eq.(\ref{eq1}) we introduce the functions $f_{\uparrow,
\downarrow}= f^{+} \pm
(\sigma_{\downarrow,\uparrow}/{\sigma_{t}})f^{-}$, where the functions
$f^{\pm}(z)$ are the $z$-dependent parts (multiplying exponential
decay factors in the $x$-direction $\exp (-kx)$) of the special
solutions of Eq.(\ref{eq1}), $\sigma_t=\sigma_{\uparrow} +
\sigma_{\downarrow}$. Due to the symmetry of the system we obtain
\begin{gather}
\label{eq2}
 f^{\pm} = \left\{
 \begin{array}{ll}
 C_N^{\pm} \ cos(\kappa_{N}^{\pm} z),          &  |z|< w/2, \\
 C_M^{\pm} \ sin(\kappa_{M}^{\pm}(d/2 - z)),   &  |z| >  w/2,\\
 \end{array} \right. \\
\kappa_{M,N}^{+}=k, \ \kappa_{M,N}^{-} = \sqrt{k^{2} -
\lambda_{M,N}^{-2}},\nonumber
\end{gather}
where the diffusion length $\lambda_{M,N}$ is the characteristic
length-scale for equilibration of the spin subsystems
in the magnetic (M) and nonmagnetic (N) regions,
respectively, $\lambda=(\sigma_0 \tau_{sf}/e^2 \Pi_0)^{1/2}$ (with
the corresponding subscripts M and N as appropriate),
$\sigma_0^{-1}=\sigma_{\uparrow}^{-1} + \sigma_{\downarrow}^{-1}$.
The functions $f_{\uparrow, \downarrow}$ have to satisfy the equations
\begin{equation}
\label{eq3} \frac{d}{dz}(\sigma_{\uparrow,\downarrow}\frac{d
f_{\uparrow,\downarrow}}{dz})=\frac{\Pi_0e^2}{\tau_{sf}}(f_{\uparrow,\downarrow}-
f_{\downarrow,\uparrow})-k^2\sigma_{\uparrow,\downarrow}f_{\uparrow,\downarrow} .
\end{equation}
Rewriting Eq.(\ref{eq3}) for the functions
$\sqrt{\sigma_{\uparrow,\downarrow}}f_{\uparrow,\downarrow}$ we
obtain an equation for the eigenfunctions of a self-adjoint
operator with the eigenvalues $k^2$.
A full set of the solutions
$\sqrt{\sigma_{\uparrow,\downarrow}}f_{{\uparrow,\downarrow}n}$ of
Eq.(\ref{eq3}) corresponding to the possible values of the
parameters $k_{n}^{2}$ is a complete basis set in the interval
$|z| < d/2$. Consequently, the general solution
of Eq.(\ref{eq1}) is given by
\begin{gather}
\label{eq4}
 \mu_{\uparrow,\downarrow}=\sum_n (a_{n}^{+} e^{k_n x}
+ a_{n}^{-} e^{-k_n
x})\frac{\sqrt{\sigma_{\uparrow,\downarrow}}f_{{\uparrow,\downarrow}n}}{\sqrt{K_n}},\\
K_n=\Bigl< (\sqrt{\sigma_{\uparrow}}f_{\uparrow n})^2 +
(\sqrt{\sigma_{\downarrow}}f_{\downarrow n})^2\Bigr>_d,
\langle...\rangle_{d,w}=\int_0^{\frac{d,w}{2}}...dz,\nonumber
\end{gather}
where the constants $a_{n}^{\pm}$ are determined by the boundary
conditions at the ends of the spin-guide, in conjunction with the
orthogonality of the functions
$(\sigma_{\uparrow,\downarrow})^{1/2}f_{{\uparrow,\downarrow}n}$
for different $n$ values.
The current densities at the channel entrance ($x=0$) are given
by $J_{\uparrow,\downarrow}(x=0, z)$, and for $\mu_{\uparrow,\downarrow}=0$
at $x=L$, where $L$ is the spin-guide length, we obtain
\begin{equation}
\label{eq5}
 a_{n}^{\pm}= \frac{ \mp c_n}{1+e^{\pm2k_nL}},
c_n=\sum_{{\uparrow,\downarrow}}\Bigl<\frac{eJ_{\uparrow,\downarrow}(x=0,
z)f_{\uparrow,\downarrow
n}}{k_n\sqrt{\sigma_{\uparrow,\downarrow}K_n}}\Bigr>_d.
\end{equation}
The coefficients $C_{N,M}^{\pm}$ (see Eq.(\ref{eq2})) are determined
by matching the functions
$\mu_{\uparrow,\downarrow}$ and the currents (i.e. the derivatives
$\sigma_{\uparrow,\downarrow}\partial\mu_{\uparrow,\downarrow}/\partial
z $) at $z = \pm w/2$ (see \cite{Fu}). The possible values of the
damping parameter $k_n$ may be found from the consistency
condition of these equations
\begin{align}
%\begin{eqnarray}
\label{eq7} &\Bigl[1- \frac{2\sigma_N}{\sigma_{Mt}}\tan
\bigl(\frac{k_n w}{2}\bigr)\tan \bigl({k_n w_M}\bigr) \Bigr]
\Bigl[ \frac{2\sigma_{M \uparrow}\sigma_{M
\downarrow}}{\sigma_{Mt}} \nonumber\\  & -\sigma_N
\frac{\kappa_{N}^{-}}{\kappa_{M}^{-}} \tan
\bigl(\frac{\kappa_{N}^{-}w}{2}\bigr)\tan \bigl({\kappa_{M}^{-}
w_M}\bigr) \Bigr] = \\ &\sigma_N \Bigl(\frac{\sigma_{M \uparrow} -
\sigma_{M \downarrow}}{\sigma_{Mt}}\Bigr)^2
\frac{k_n}{\kappa_{M}^{-}} \tan \bigl(\frac{k_n w}{2}\bigr)
\tan\bigl({\kappa_{M}^{-} w_M}\bigr).\nonumber
%\sigma_{Mt}=\sigma_{M \uparrow} + \sigma_{M \downarrow}.\nonumber
\end{align}
If constant current is injected into the NM
channel only (i.e. at $x=0$:  $I_{\uparrow,\downarrow}= I_0/2$ at
$z<w/2$ and $I_{\uparrow,\downarrow}=0$ at $z > w/2$), and
restricting ourselves to current variation only in the $x$
direction (i.e. averaging in the transverse, $z$, direction),
the current in the channel is given by
\begin{equation}
\label{eq6} I_{\uparrow, \downarrow}=\frac{2I_0}{w}
\sum_{n}\frac{\sigma_N\cosh(k_n(L-x))}{K_n\cosh(k_n
L)}\Bigl(\langle f_{n}^{+}\rangle^2_w \pm \frac{\langle
f_{n}^{+}\rangle_w \langle f_{n}^{-}\rangle_w }{2}\Bigr).
\end{equation}
From the above we can calculate the degree of spin-polarization of
the current in the channel, $\alpha =
(I_{\uparrow}-I_{\downarrow})/(I_{\uparrow}+I_{\downarrow})$. The
solutions given in Eq.(\ref{eq6}) exhibit exponential decreases
along the channel, with different decay rates for the two spin
directions.
The main contribution to the spin-up current is
$I_{\uparrow}(x)\propto exp(-k_{\uparrow}x)$ where
$k_{\uparrow}\approx k_1\equiv k_{min}$ is the smallest solution
from the set $\{k_n\}$ determined by Eq.(\ref{eq7}), while the
spin-down current decays much faster $I_{\downarrow}(x)\propto
exp(-k_{\downarrow}x)$, where $k_{\downarrow}\approx k_2 >
k_{\uparrow}$. Thus, inside the spin-guide the spin polarization
of the current in the nonmagnetic channel will tend exponentially
to unity, i.e. $\alpha \approx
1-exp(-(k_{\downarrow}-k_{\uparrow})x)$; for further details see
{\cite{Fu}}.
\begin{figure}
\includegraphics[height=5cm,width=8cm]{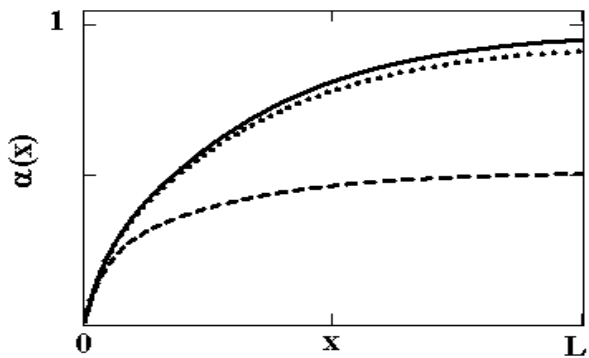}
\caption{The degree of spin-polarization ($\alpha $) of the
current plotted vs. the longitudinal ($x$) coordinate along the
spin guide; the curves were calculated from Eq.(\ref{eq6}) with
$\sigma_{M\uparrow}/\sigma_{M\downarrow} = 0.3$,
$\sigma_{M\downarrow}/\sigma_{N} = 1$, $w/d = 0.28$, $L = 4d$,
$w_M/\lambda_M=0.225$, and $w/\lambda_N = 0.1$ (solid),
$w/\lambda_N = 0.5$ (dotted), $w/\lambda_N = 0.7$ (dashed).}
\label{fig2}
\end{figure}

In Fig.{\ref{fig2}} we display the $x$-dependence of the degree of
spin-polarization calculated from Eq.(\ref{eq6}).
These curves demonstrate exponential growth of the degree of
spin-polarization  with distance from the entrance. In general,
the saturation value is dependent on spin-flip processes in the
materials used for the device. Comparison of the curves for
different $w/\lambda_N$ illustrates that higher current
polarization is achieved  for smaller ratios between the width of
the NM channel and the spin-flip length $\lambda_N$.
The current and spin polarization in the device may be manipulated
and tuned continuously with an applied magnetic field. Indeed, by
changing the magnetic field the ratio $\sigma_{M\uparrow}$ /
$\sigma_{M\downarrow}$ can be changed, and one may tune the
current decay parameter k (see Eq.(\ref{eq7})).

\begin{figure}
\includegraphics[height=5cm,width=8cm]{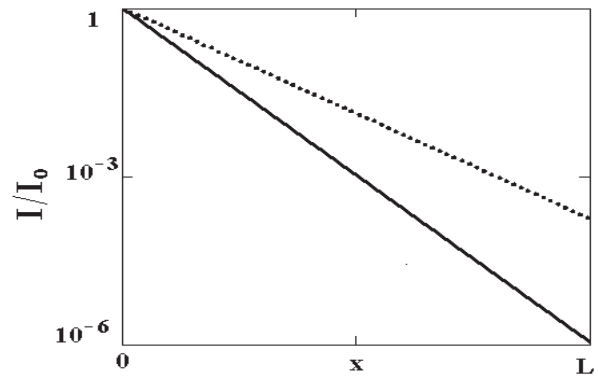}
\caption{The normalized current $I/I_0$ (logarithmic scale)
plotted vs. the longitudinal ($x$) coordinate along the spin
guide. Eq.(\ref{eq6}) was used, for magnetized (dotted, exhibiting
a high value at $x=L$) and unmagnetized (solid) shells, with the
parameters of the solid line in Fig.{\ref{fig2}}. } \label{fig3}
\end{figure}

Switching-off of the magnetic field increases the damping factor
of the current from the value $k_{min}$ (in the case of an ideal
magnetic material shell $k_{min} \approx (\lambda_N
\sqrt{2})^{-1}$ {\cite{Fu}}) up to $k_{UM}$ which depends on the
ratio of the conductivities of the unmagnetized (UM) shell and the
NM channel:
$\tan\bigl(k_{UM}w/2\bigr)\tan\bigl(k_{UM}(d-w)/2)\bigr)=
\sigma_{UM}/\sigma_N$;  for the present case this equation
replaces Eq.(\ref{eq7}). When $\sigma_{UM} \approx \sigma_N $ we
obtain $k_{UM}\approx \pi/d$. Fig.{\ref{fig3}} depicts the
variation of the current along the nonmagnetic channel when a
magnetized (dotted line) or an unmagnetized (solid line) shell is
used - the current changes by about
three orders of magnitude at the exit ($x = L$) of the MFET.
When $\sigma_{M\uparrow,\downarrow} \gg \sigma_N $ the difference
between $k_{min}$ and $k_{UM}$ is small and the field effect is
also small.
Increase of the NN/MM interface resistance (e.g. via adding
barriers), will enhance the effect.

A giant negative magnetoresistance effect may be observed by
switching the magnetization directions in the upper and lower
magnetic layers from being parallel to each other, to having
antiparallel directions. When the upper and lower magnetic layers
have parallel magnetization directions, a polarized current will
arrive at the channel exit, since electrons with only one spin
direction will be transported preferentially  through the magnetic
layers to the grounding. In contrast, when the magnetic layers'
magnetizations are antiparallel, the output current will be
unpolarized and it will decrease significantly in magnitude.
This effect could be of particular interest in the case of
a ferromagnetic metal shell where residual magnetization may
remain (because of domain structure) upon switching-off the
magnetic field.

%This originates from the fact that while on/off magnetic field
%switching may be operative for a DMS (where complete
%demagnetization may be expected upon switching-off the magnetic
%field), it may fail when a ferromagnetic metal is used as the
%magnetic shell since a certain  degree of spin selectivity may
%remain upon switch-off of the magnetic field (because of domain
%structure). Therefore, in the latter case, the
%parallel/antiparallel changes in the magnetization directions may
%be preferred.

In summary, we proposed a scheme for a magnetic-field-effect
transistor, i.e. a device for generation of
highly spin-polarized currents, whose operation is governed by a
magnetic field.
In this device nearly complete current spin-polarization
may be achieved, even when a magnetic shell with a
lower degree of spin polarization is used.
Large changes in the current may be brought about through on/off
switching of a magnetic field.
Furthermore, the magnetic-field sensitivity suggests the MFET as
a magnetic-field-sensor.
Finally, the wide range of measured spin-diffusion lengths
(typically of the order of $10$ nm in metals and microns in
semiconductors {\cite{To}}) suggests the potential fabrication of
nano-scale MFET devices, with possible incorporation of known
nanostructures (e.g. nanotubes as the conducting channel).

This research was supported by Grant No. UP2-2430-KH-02 of the
CRDF and by the US DOE Grant No. FG05-86ER-45234 (E.N.B and U.L.).
%\begin{thebibliography}{99}
%\begin{bibliography}
\begin{itemize}
\bibitem{Aws} {\it Semiconductor Spintronics and Quantum
Computation}, eds. D. D. Awschalom, D. Loss, and N. Samarth,
Springer, Berlin, 2002.
\bibitem{Dat} S. Datta and B. Das, Appl. Phys. Lett. {\bf 56}, 665 (1990).
\bibitem{Aro} A.G. Aronov, JETP Lett. {\bf 24}, 32 (1976).
\bibitem{Son} P. C. van Son, H. van Kempen, and P. Wyder Phys. Rev. Lett. {\bf 58},
2271 (1987).
\bibitem{Mol2} G. Schmidt, G. Richter, P. Grabs, C. Gould, D. Ferrand, and L.
W. Molenkamp, Phys. Rev. Lett. {\bf 87}, 227203 (2001).
\bibitem{Fu} R.N.Gurzhi, A.N.Kalinenko, A.I.Kopeliovich, A.V.Yanovsky,
E.N.Bogachek, and Uzi Landman, Phys. Rev. B 68, 125113 (2003).
\bibitem{To}  {\it Spin electronics}, eds. M. Ziese, M.J.Thornton,
Springer, Berlin; New York, 2001.
\end{itemize}
%\end{thebibliography}%1

\newpage

\end{document}